\documentclass[showpacs, pra,twocolumn,preprintnumbers ,amsmath, amssymb, superscriptaddress, aps]{revtex4-2}

\DeclareUnicodeCharacter{202F}{\,}
\usepackage{inputenc}
\usepackage{color}
\usepackage{amsmath,amssymb}
\usepackage{pifont}
\usepackage{amssymb}  
\usepackage{bbold}
\usepackage{float}
\usepackage{subfloat}

\usepackage[caption=false]{subfig}
\usepackage{tikz}
\usepackage{makecell}
\usepackage{subfig}
\usepackage{pifont}   
\usepackage{graphicx} 
\graphicspath{{Figures4/}}
\usepackage{dcolumn}  
\usepackage{bm}       
\usepackage{multirow} 
\usepackage{placeins}
\usepackage[colorlinks]{hyperref}
\usepackage{mathtools}
\usepackage{appendix}

\def \be{\begin{align}}
	\def \ee{\end{align}}
\def \bea{\begin{eqnarray}}
	\def \eea{\end{eqnarray}}

\begin{document}
\title{Spin and valley-dependent tunneling in  MoS$_2$ through  magnetic barrier}
	

\author{Ahmed Jellal}
\email{a.jellal@ucd.ac.ma}
\affiliation{Laboratory of Theoretical Physics, Faculty of Sciences, Choua\"ib Doukkali University, PO Box 20, 24000 El Jadida, Morocco}
\author{Nadia Benlakhouy}
\affiliation{School of Applied and Engineering Physics, University Mohammed VI Polytechnic, Ben Guerir, 43150, Morocco}
\author{Pablo Díaz}
\affiliation{Departamento de Ciencias F\'{i}sicas, Universidad de La Frontera, Casilla 54-D, Temuco 4811230, Chile}  
\author{David Laroze}
\affiliation{Instituto de Alta Investigación, Universidad de Tarapacá, Casilla 7D, Arica, Chile}

\begin{abstract}

We study electron transport in monolayer molybdenum disulfide MoS$_2$ subjected to a magnetic barrier. Our analysis employs a full-band continuum model to capture the relevant physical phenomena. We focus on how electron energy, magnetic field strength, and the geometric characteristics of the barrier affect the transmission and conductance. We observe sharp resonant tunneling features emerging from quantum interference effects induced by magnetic confinement. A key outcome of our study is the discovery of distinct resonance patterns in the conduction and valence bands. These patterns are closely related to the intrinsic spin-orbit coupling in MoS$_2$ and the breaking of time-reversal symmetry by the magnetic field. This results in significant spin and valley selectivity in electron transport. We demonstrate that adjusting external parameters precisely controls spin-polarized and valley-polarized currents. We show that a magnetic barrier can control electron spin and valley in MoS$_2$,
making it a promising platform for energy-efficient spintronic and valleytronic devices.

	\end{abstract}
	\pacs{}
	\maketitle
	\section{Introduction}

Two-dimensional (2D) materials exhibit unique properties for applications in nanoelectronics, sensing, and photonics~\cite{xia2014two,liu2019van}. Besides graphene, transition metal dichalcogenides (TMDs) such as MX$_2$ (M = Mo, W; X = S, Se, Te) have attracted attention~\cite{chowdhury2020progress,fu20212d,yang20232d,yun2020layered}. Among them, the monolayer MoS$_2$ is particularly promising due to its direct bandgap, strong spin-orbit coupling, and compatibility with semiconductor fabrication. These properties enable electronic~\cite{lv2015transition,gatensby2014controlled,lembke2015single}, optoelectronic~\cite{gong2017electronic,pospischil2016optoelectronic,choi2017recent}, and spintronic~\cite{ahn20202d,feng2017prospects,li2016two} applications. Unlike graphene~\cite{geim2007rise,geim2009graphene,cooper2012experimental}, MoS$_2$ couples spin and valley properties~\cite{splendiani2010emerging,mak2010atomically,xiao2012coupled}, allowing control of carriers by charge, spin, and valley index. This leads to novel quantum effects, including unconventional quantum Hall states and combined spin/valley Hall effects~\cite{xiao2012coupled}, and advanced devices, such as high-performance transistors and spin-valleytronic systems~\cite{wang2012large,wang2012integrated}. In addition, MoS$_2$ can be mass-produced by chemical vapor deposition~\cite{lee2012synthesis,liu2012growth,zhan2011large} and integrated with other 2D or flexible materials~\cite{wang2012integrated}.

 The charge transport properties of MoS$_2$ are critical in determining device performance. Different studies have investigated the carrier dynamics in MoS$_2$, analyzing both intrinsic material properties and external influences such as defects, mechanical strain, and applied fields \cite{radisavljevic2011single, kaasbjerg2012phonon}. A distinctive feature of MoS$_2$ is its spin-valley coupling, which leads to unique transport properties that are absent in gapless systems such as graphene \cite{xiao2012coupled, mak2010atomically}.
 Many studies have investigated how magnetic barriers affect the transport properties of 2D materials. In graphene, these barriers can cause Fabry-Pérot resonances. These occur when forward and backward electron waves interfere with each other, creating distinct patterns in how electrons pass through or reflect \cite{cheianov2006selective, katsnelson2006chiral}.  
 MoS$_2$ behaves differently because it has a band gap and strong spin-orbit coupling. These features provide additional control through spin and valley effects. Research shows that magnetic fields in MoS$_2$ can lead to spin- and valley-specific transport, such as resonance tunneling and polarization. These effects do not occur in bandgapless materials.  
 These results suggest that magnetic barriers can control how electrons move by changing the way they interfere. This could help us better control spin and valley behavior in future devices \cite{mak2019probing, premasiri2019tuning}.

We examine how electrons move through a monolayer of molybdenum disulfide MoS$_2$ exposed to a magnetic barrier. To capture the essential physics, we use a full-band continuum model that takes into account the material's unique electronic structure. Our main focus is on how the transport properties—specifically, transmission and conductance—respond to changes in electron energy, the strength of the magnetic field, and the size and shape of the barrier. We find that when electrons pass through the magnetic barrier, they can form quasi-bound states due to quantum interference, which gives rise to sharp resonance peaks in the transmission. These resonances are different in the conduction and valence bands, and they are strongly influenced by spin-orbit coupling, a key property of MoS$_2$. The magnetic field also breaks time-reversal symmetry, which plays a major role in enhancing the differences between electron behavior in the two valleys ($K$ and $K'$). This leads to a high degree of selectivity in both spin and valley transport. By tuning external parameters like the magnetic field or barrier geometry, we can control whether the current is spin-polarized or valley-polarized. These results suggest that magnetic barriers offer an effective way to manipulate electron flow in two-dimensional materials.

The paper is organized as follows. In Sec. \ref{S1}, we present the theoretical framework using a continuum model to describe the electronic properties of MoS$_2$. We derive the energy spectrum for the three regions of the system.  In Sec. \ref{TTRR}, we calculate the transmission and reflection probabilities by applying continuity conditions to the eigenspinors at the interfaces $x = 0$ and $x = L$. We then present a discussion based on numerical results for different choices of physical parameters.  valleys $(K, K')$. In Sec. \ref{CCCC}, we compute and analyze the conductance for both valleys $(K, K')$.
In Sec. \ref{MoGr}, we present a systematic comparison between MoS$_2$ and graphene, analyzing both their similarities and fundamental differences.
 Finally, Sec. \ref{S4} provides a summary and conclusion of our results.
	\label{Intro}

	\section{Energy spectrum}	\label{S1}

\begin{figure}[ht]
	\centering 
	\includegraphics[scale=1.2]{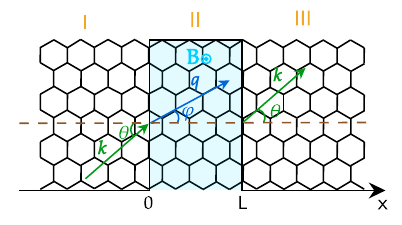}
	\caption{Schematic of a monolayer MoS$_2$ sheet with three regions (I, II, III) and a magnetic barrier of width $L$ placed in region II.  An electron incident on region I at an angle $\theta$ passes through region II at a refraction angle $\varphi$ and emerges in region III. 
		}\label{barrier}
\end{figure}

We start by defining our system, which consists of three regions of monolayer MoS$_2$. The middle region is subjected to a magnetic barrier. This barrier is modeled as two closely spaced delta-function-like magnetic fields, as shown in Fig. \ref{barrier}. It is generated by a magnetic field $B$ applied along the $z$-direction, which is
\begin{equation}
B(x) = Bl_B \left[\delta(x) - \delta(x - L)\right],
\end{equation}
where $B$ is the magnetic field strength, $l_B = \sqrt{\frac{\hbar}{eB}}$ is the magnetic length, and $L$ is the distance between the two magnetic barriers. This setup effectively models how a magnetic barrier can be obtained. In practice, such a configuration can be created by placing a ferromagnetic strip magnetized along the $x$-axis on top of the MoS$_2$ layer. Similar methods have been used in graphene experiments \cite{nogaret2000resistance, papp2001spin}. Since $B_z(x)$ does not vary along the $y$-direction, the associated magnetic vector potential takes a step-like form. Specifically, it is given by
\begin{equation}
A_y(x) = B\ell_B \left\{\theta(x) - \theta(x - L)\right\},
\end{equation}
where $\theta(x)$ is the Heaviside step function. 

In the low-energy model, the Hamiltonian describing electron motion in MoS$_2$ near the $K$ and $K'$ points under a magnetic field can be written as 
\begin{equation}
H =  v \left[ \tau \sigma_x p_x + \sigma_y \left(p_y+{e }A_y(x)\right) \right] + \frac{\Delta}{2}\sigma_z+ \frac{1-\sigma_z}{2}\lambda \tau s_z ,
\end{equation}
where $\tau = \pm$ represents the $K$ and $K'$ valleys, and $p_x$ and $p_y$ are the momentum components at these points. The Pauli matrices ($\sigma_x, \sigma_y, \sigma_z$) describe the coupling between the conduction and valence bands. The mass term $\Delta = 1.8\,\text{eV}$ breaks the inversion symmetry, while the spin splitting $\lambda = 0.082 \,\text{eV}$, caused by spin-orbit coupling at the valence band edge, further modifies the band structure \cite{li2013unconventional, zahid2013generic}. The Fermi level is given by $v = \frac{at}{\hbar} \approx 0.53 \times 10^6\,\text{m/s}$, where $t$ is the effective hopping parameter between the Mo $d$-orbitals and the S $p$-orbitals. 

Next, we determine the eigenspinors and energies in each region. In fact, due to the translational invariance along the $y$-direction, we can decompose the spinor as
\begin{equation}
\psi(x, y)= \phi (x){e}^{\mathrm{i} k_y \cdot y},
\end{equation}
and to explicitly determine $\phi (x)$, we solve the eigenvalue equation $H\psi=E\psi$.  In region I,  we find the solution
\begin{align}
 \phi_\text{I}(x)=\binom{1}{ e^{i \theta}} e^{ik_x x}+r\binom{1}{ -e^{-i \theta}} e^{-ik_x x} , 
\end{align}
where  $\theta=\tan ^{-1}\frac{k_y} {k_x}$ is the incident angle  and $r$ is the reflection coefficient.  
 The corresponding energy is given by
\begin{equation}\label{ee1}
E_{1s}= \frac{\lambda\tau s_z}{2} +s  \hbar v_F\sqrt{k^2+\left(\frac{\lambda\tau s_z -\Delta}{2  \hbar v_F }\right)^2},
\end{equation}
with  $s=\pm$ and $k=\sqrt{k_x^2+k_y^2}$. We can  derive  the wave vector component $k_x$ as
\begin{align}
k_x =\tau\sqrt{\frac{1}{\hbar^2 v_F^2}\left(E + \frac{\Delta}{2} \right)  \left(E- \lambda\tau s_z+ \frac{\Delta}{2} \right)- k_y^2}.
\end{align}
 In region III, we show that $\phi(x)$ has the form
 \begin{align}
 \phi_\text{III}(x)=t\binom{1}{e^{i \theta}} e^{ik_x x},
\end{align}
 where $t$ is the transmission coefficient.
 In region II, the solution is 
\begin{align}
 \phi_\text{II}(x)=a\binom{1}{ e^{i \varphi}} e^{iq_x x}+b\binom{1}{ - e^{-i \varphi}} e^{-iq_x x},
\end{align}
with the angle $\varphi=\tan ^{-1}\left(\frac{k_y+l_B^{-1}} { q_x}\right)$, and 
the associated energy is
\begin{equation}\label{ee2}
E_{2s'}= \frac{\lambda\tau s_z}{2} + s' \hbar v_F  \sqrt{q_x^2+ \left(k_y+l_B^{-1}\right)^2+\left(\frac{\lambda\tau s_z -\Delta}{2v_F \hbar }\right)^2},
\end{equation}
from which we can obtain  $q_x$ as
\begin{equation}
q_x=\tau \sqrt{\frac{\left(E+\frac{\Delta}{2}\right)\left(E- \lambda\tau s_z+ \frac{\Delta}{2} \right)}{\hbar^2v_F^2} -\left(k_y+l_B^{-1}\right)^2},
\end{equation}
with the sign $s'=\pm$.

\begin{figure}[ht!]
	\centering 
	\includegraphics[scale=0.5]{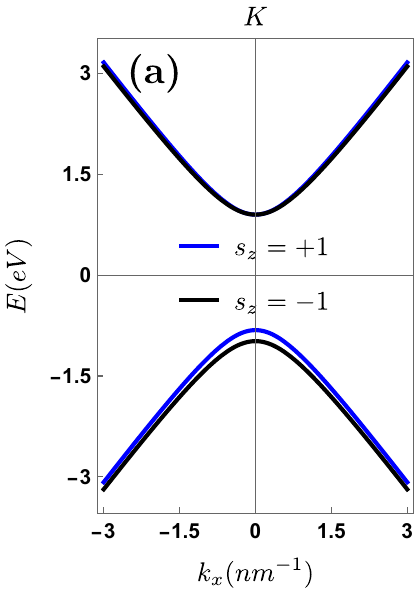}
	\includegraphics[scale=0.5]{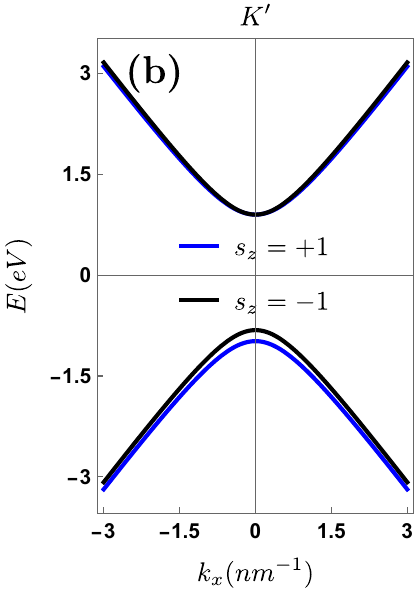}\\ 
	\includegraphics[scale=0.5]{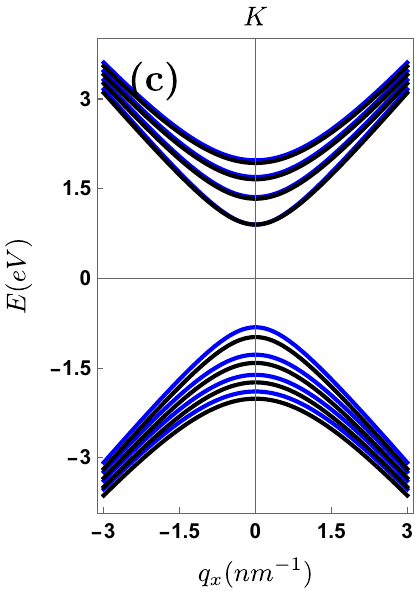}
	\includegraphics[scale=0.5]{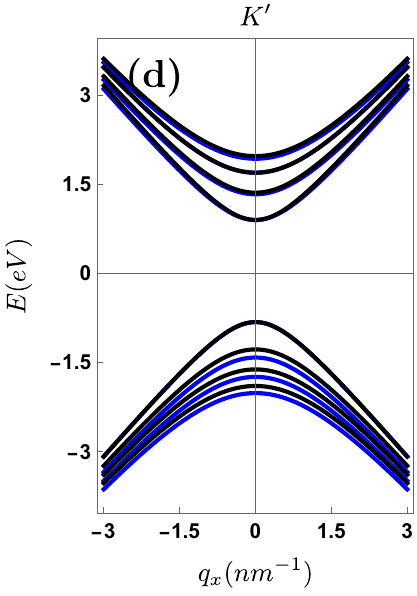}
	\label{Energy}
	\caption{
		Energies \eqref{ee1} and \eqref{ee2} as a function of $k_x$ and $q_x$ for  $B=0, 1, 2, 3$ T. Panels (a) and (b) display the energy dispersion in the normal regions (region I and region III, respectively), where no magnetic field is present. Panels (c) and (d) illustrate the modified band structure in region II under the influence of the magnetic barrier. }
		\label{Energy}
\end{figure}

Fig. \ref{Energy} shows the energy spectra   \eqref{ee1} and \eqref{ee2} versus the wave vector components $k_x$ and $q_x$.
 We choose $B= 0-3$~T to show how the system changes from spin-orbit effects at low fields to clear Landau levels at high fields, where spin and valley splitting becomes visible.
In regions I and III, the band structure of  MoS$_2$ is clearly visible. It is characterized by a large direct bandgap and a spin-orbit coupling splitting. The upper valence bands originate mainly from Mo $d_{x^2-y^2}$ and $d_{xy}$ orbitals, while the lower conduction bands originate mainly from Mo $d_{z^2}$ orbitals. In region II, which is affected by a magnetic barrier, the band structure is modified. This change is due to the vector potential, which shifts the transverse momentum and changes the band splitting. As $B$ increases  ($0\to 3$) T, the band structure changes become more pronounced. 
Mini-gaps form, and valley effects become more visible and easier to control \cite{cheng2015transport}.

\section{Transmission}\label{TTRR}
The coefficients $r, a, b$, and $t$ can be determined from the continuity of  eigenspinors at the interfaces $x=0, L$. 
This process allows us to write 
\begin{equation}
    \phi_\text{I}(0)=\phi_\text{II}(0),\quad
    \phi_\text{II}(L)=\phi_\text{III}(L),
\end{equation}
which gives the set of equations
\begin{align}
	& 1+r=a+b ,\\
	& e^{i \theta}-r  e^{-i \theta}=a  e^{i \varphi}-b  e^{-i \varphi}  ,\\
	& a e^{i q_x L}+be^{-i q_x L}=t e^{i k_x L}, \\
	& a e^{i \varphi}  e^{i q_{x}L}-b  e^{-i \varphi} e^{-i q_x L}=t e^{i \theta}\ e^{ik_x L}.
\end{align}
After some algebras, we get  the transmission and reflection coefficients
\begin{widetext}
\begin{align}
& t = \frac{e^{-ik_x L} \cos\theta \cos\varphi}{   \cos( q_xL) \cos\theta \cos\varphi-i\sin( q_xL)\left(1- \sin\theta \sin\varphi\right)},
\\
&r = \frac{4 e^{i( q_xL +2 \theta+\varphi)} \sin( q_xL)\left( \sin \theta- \sin\varphi\right)}{e^{2 i \varphi}\left(e^{2 i  q_xL}+e^{2 i \theta}\right) +\left(1+e^{2 i( q_xL+\theta)}\right) -2e^{i(\theta+\varphi)}\left(-1+e^{2 i  q _xL}\right)}.
\end{align}
\end{widetext}
A straightforward calculation results in the transmission  probability
\begin{equation}\label{T}
T= \frac{ \cos^2 \theta\cos^2 \varphi}{ \cos^2(q_xL ) \cos^2  \theta \cos^2 \varphi+\sin^2(q_xL )\left(1- \sin\theta \sin\varphi\right)^2}.
\end{equation}
To explore the main features of our system, we perform a detailed numerical study of the transmission behavior. By analyzing the transmission, we aim to show how the system responds to changes in magnetic field strength $B$ and barrier width $L$. Our results help to explain the resonant transmission effects and provide a clearer view of the mechanisms that control transport under such complex conditions.
We selected the 0.1-0.4 T magnetic field range in our plots for two key reasons. First, this range ensures the magnetic length remains comparable to our system size. Second, it preserves the important balance between spin-orbit and magnetic confinement effects. At higher fields, these effects would be obscured by dominant Landau quantization.

 \begin{figure}[htp!]
\centering 
\includegraphics[scale=0.35]{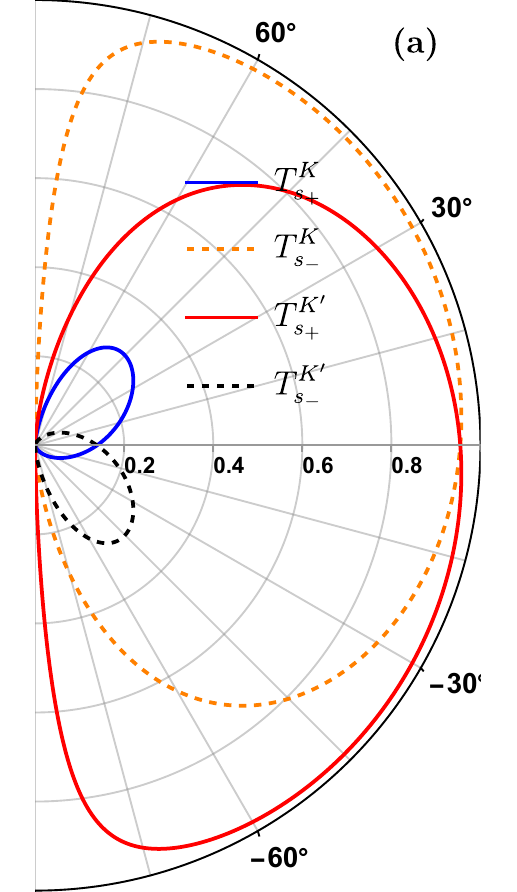}~~~~~~~~\includegraphics[scale=0.35]{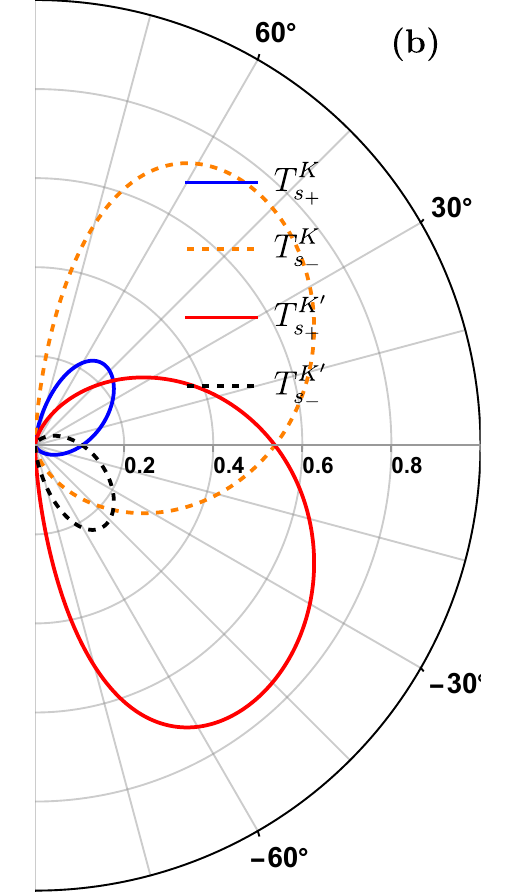}\\
\vspace{0.21cm}
\includegraphics[scale=0.35]{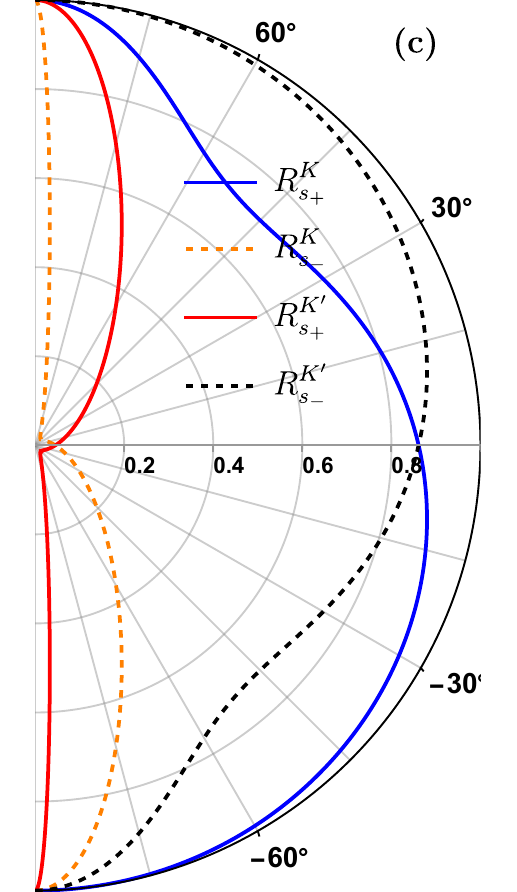}~~~~~~~~\includegraphics[scale=0.35]{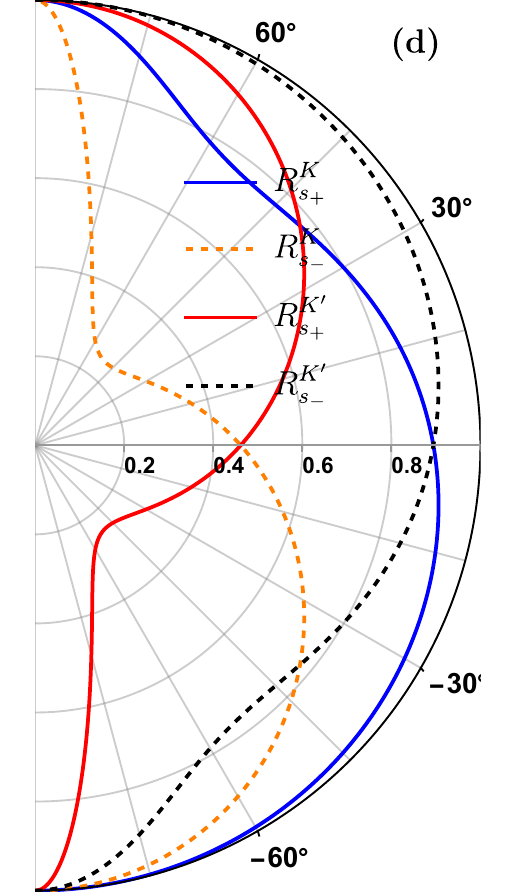}
\caption{Transmission and reflection probabilities versus the incident angle $\theta$ for $E=2.5$ eV, $k_y=0.5~\text{nm}^{-1}$,  $L=15$ nm, and two magnetic field values  (a,c):  $B=0.1$ T,  (b,d): $B=0.2$ T. }\label{T0}
\end{figure}

Fig. \ref{T0} shows how electrons in MoS$_2$ are transmitted and reflected at different angles under a magnetic barrier. In (a) and (c) the magnetic field is 0.1 T, while in (b) and (d) it is 0.2 T. The energy is 2.5 eV, and the barrier width is 15 nm. At $B=0.1$ T, both K and $K'$ valleys show strong transmission at normal incidence. For $B=0.2$ T, we observe that transmission decreases and reflection increases. The valleys also behave differently, showing a clear asymmetry. This happens because MoS$_2$ has strong spin-orbit coupling, which affects the four spin-valley channels differently. We conclude that varying the magnetic field helps to control the spin- and valley-polarized electron flow in MoS$_2$. These results are consistent with previous studies \cite{xiao2012coupled,mak2014valley}.

\begin{figure}[htp!]
\centering 
\includegraphics[scale=0.4]{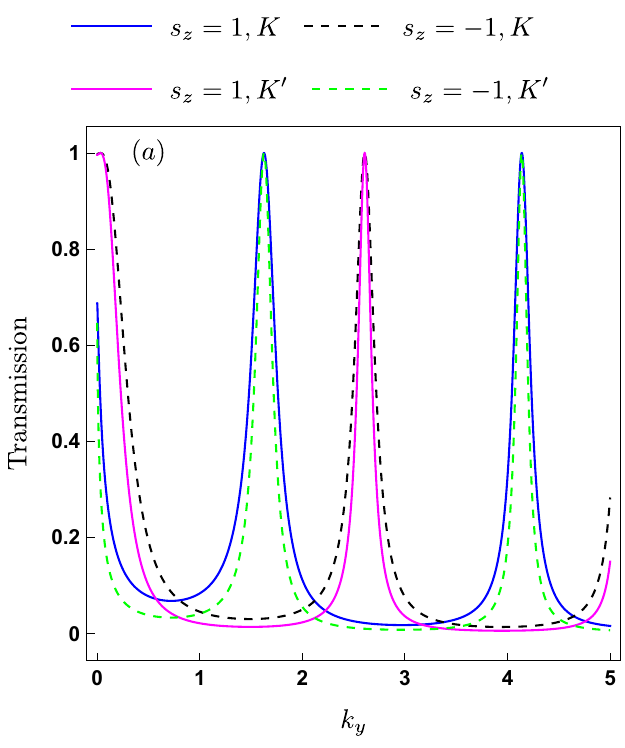}	\includegraphics[scale=0.4]{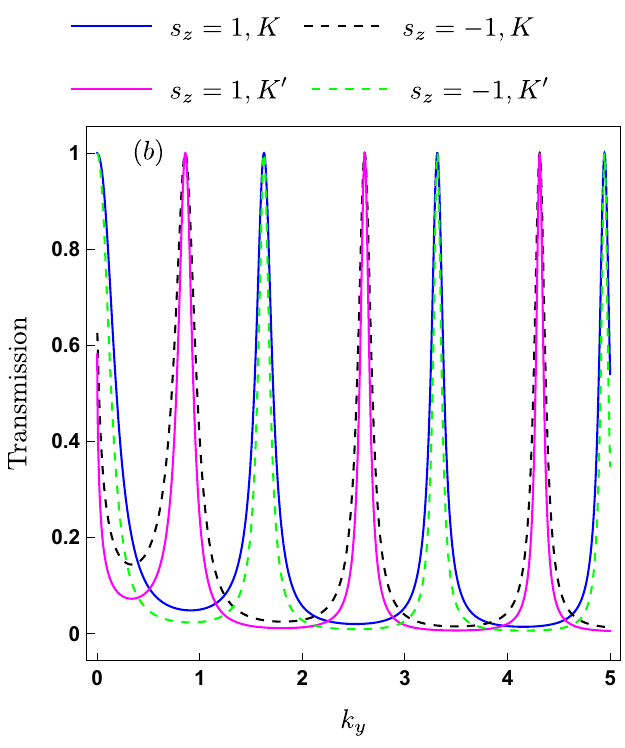}\\
\includegraphics[scale=0.4]{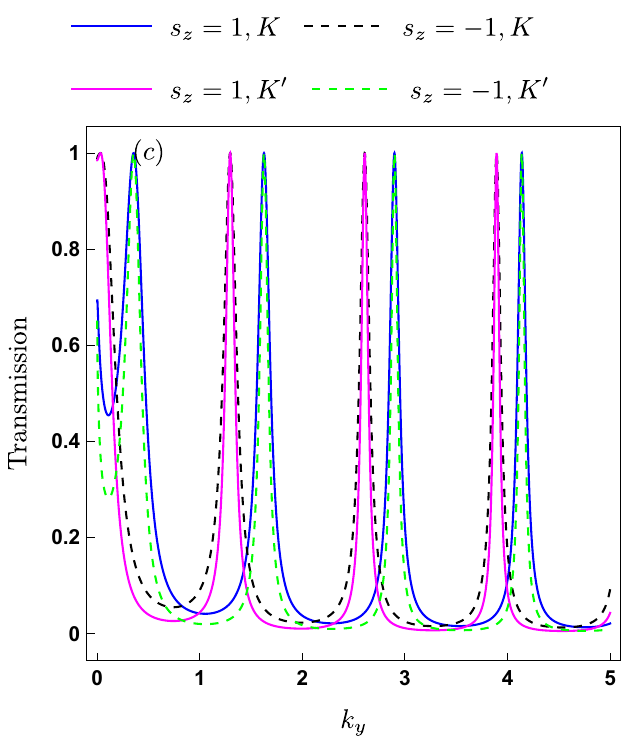}\includegraphics[scale=0.4]{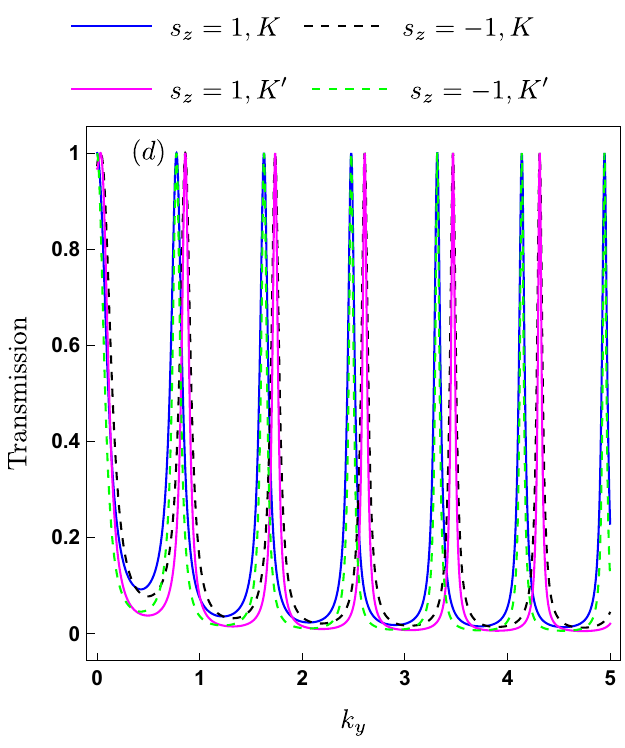}
\caption{Spin and valley resolved transmission probability for electrons in the $K$ and $K'$ valley as a function of the transverse momentum $k_y$ for $E=2.5$ eV,   $B=0.1$ T, and four barrier width values
(a): $L=10$ nm, (b): $L=15$ nm, (c): $L=20$ nm,  (d): $L=30$ nm.}\label{T1}
	\end{figure}
  
\begin{figure}[htp!]
\centering 
\includegraphics[scale=0.4]{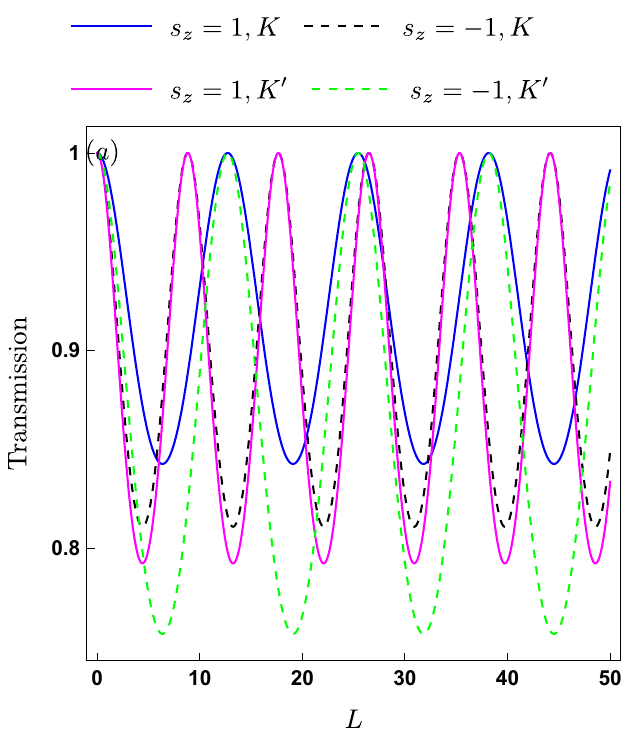}	\includegraphics[scale=0.4]{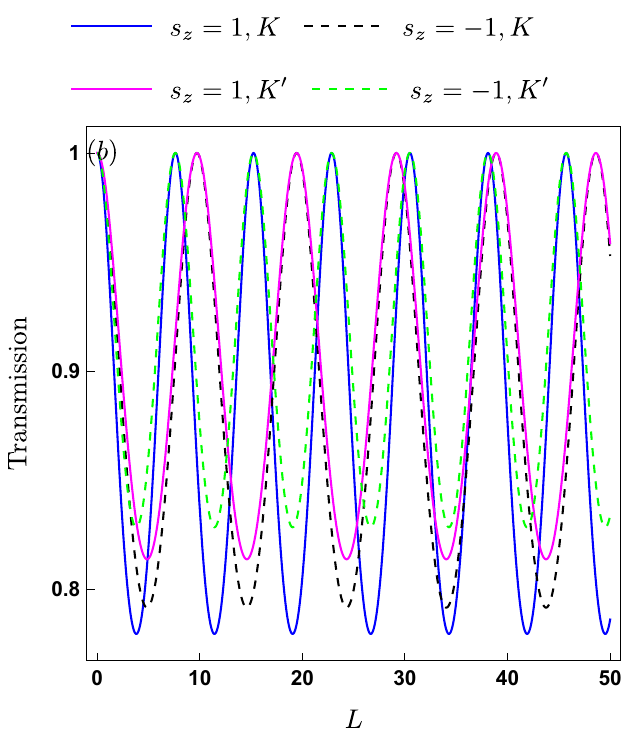}\\
\includegraphics[scale=0.4]{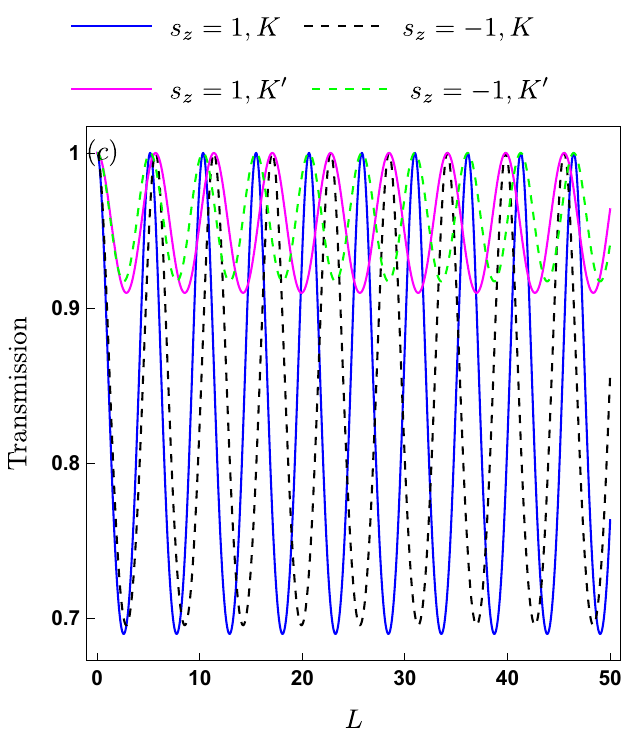}\includegraphics[scale=0.4]{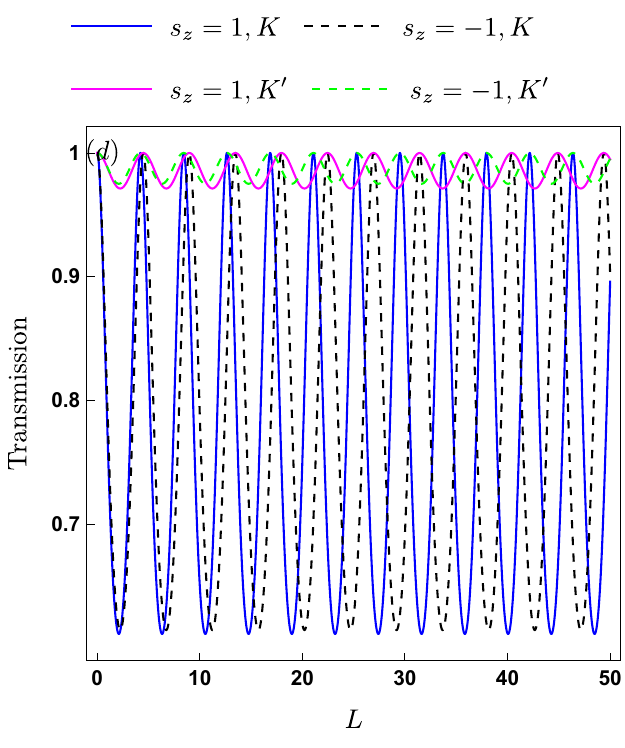}
\caption{Transmission probability versus the magnetic barrier width $L$ in $K$ and $K'$ valleys for $E=2.5$ eV,   $k_y=0.5~\text{nm}^{-1}$, and four magnetic field values  (a): $B= 0.1$ T,   (b):  $B=0.2$ T, (c): $B=0.3$ T, (d):   $B= 0.4$ T.}\label{T2}
\end{figure}

Fig. \ref{T1} shows the spin- and valley-dependent transmission probabilities versus the transverse momentum $k_y$ for different barrier widths. For a narrow barrier ($L=10$ nm, Fig. \ref{T1}a), both valleys show similar oscillation patterns, resulting in weak spin filtering in the $K$-valley. As the barrier widens (Fig. \ref{T1}b-d), the oscillations become more frequent and their amplitudes change, indicating stronger quantum confinement effects. The $K$ and $K'$ valleys develop opposite transmission behaviors—when one valley enhances transmission for a particular spin, the other suppresses it. This complementary effect allows selective spin and valley filtering by tuning the barrier width. In particular, certain $k_y$ regions produce sharp transmission peaks from quantum interference that are extremely sensitive to small variations in $L$. Such sensitivity suggests that small adjustments to system parameters could effectively control spin and valley polarization, highlighting potential applications in nanoelectronics.

Fig.~\ref{T2} shows how the transmission probability changes with the magnetic barrier width $L$ for electrons in the $K$ and $K'$ valleys for  $E = 2.5$ eV, $k_y = 0.5~\text{nm}^{-1}$, and four magnetic field values $B = 0.1, 0.2, 0.3, 0.4$ T in (a), (b), (c),  (d),   respectively. The transmission behavior results from the interference between forward and backward electron waves inside the barrier. This interference is described by the oscillatory terms $\cos(q_xL)$ and $\sin(q_xL)$ in~\ref{T}. When the condition $q_xL = n\pi$ is satisfied, constructive interference occurs, leading to resonance peaks in the transmission probability. For $B = 0.1$ T, both valleys show clear oscillations, with resonance peaks becoming more pronounced as $L$ increases. For $B=0.2$ T, the effective wave vector $q_x$ becomes larger because the magnetic confinement is stronger. This shifts the resonance conditions and changes the frequency and strength of the oscillations. At higher magnetic fields ($B = 0.3, 0.4$ T), a clear valley-dependent behavior emerges. The $K$ valley retains sharp, regularly spaced resonance peaks, while the $K'$ valley exhibits either weakened or shifted peaks. This difference is due to the strong spin-orbit coupling of MoS$_2$, which modifies the effective potential and quantum interference conditions differently in each valley.

\begin{figure}[htp!]
\includegraphics[scale=0.4]{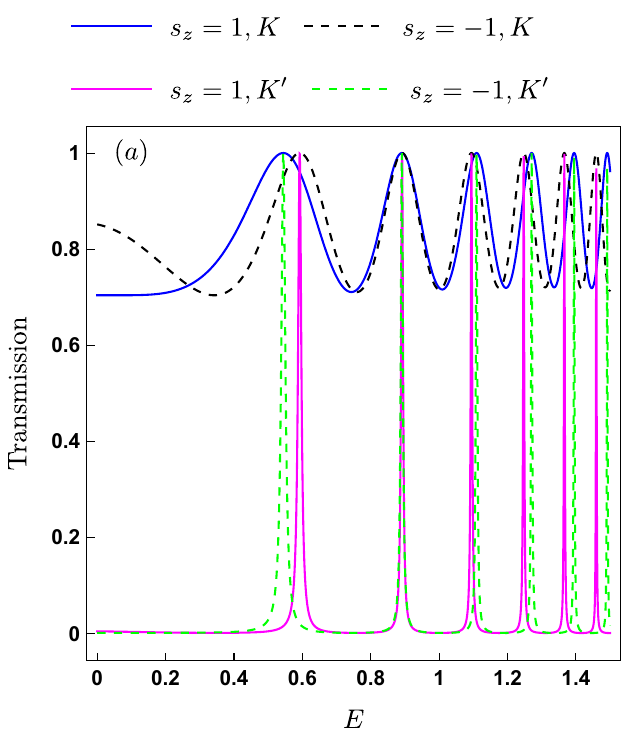}	\includegraphics[scale=0.4]{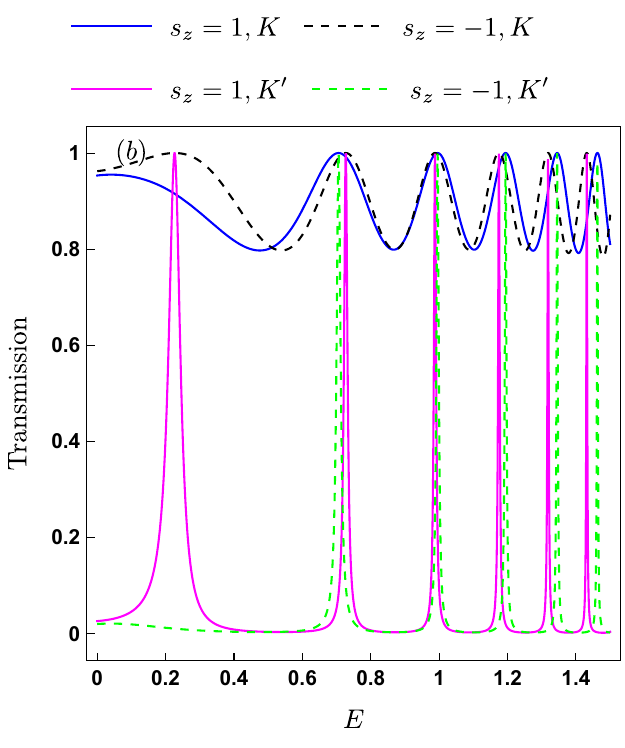}\\
\includegraphics[scale=0.4]{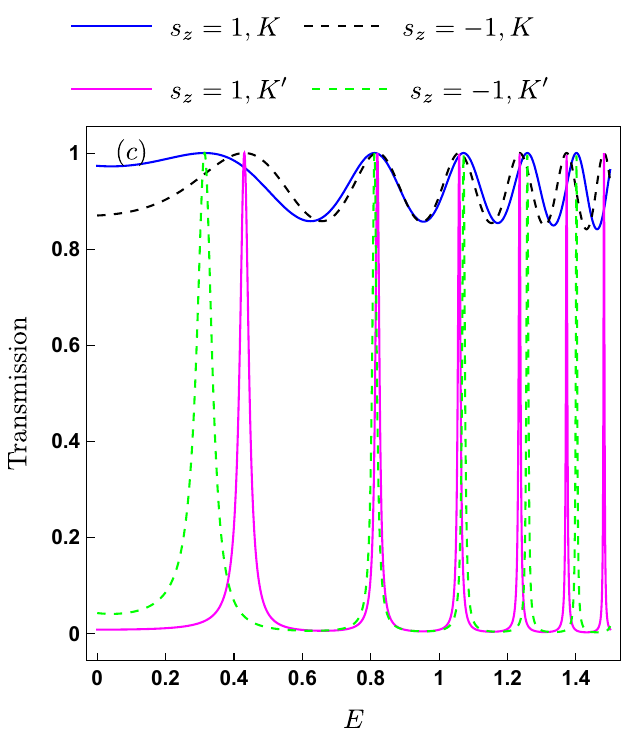}\includegraphics[scale=0.4]{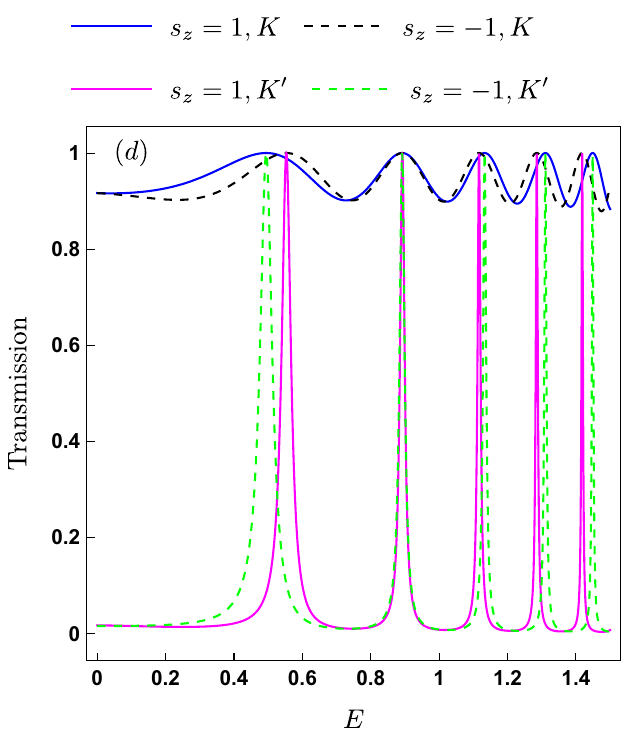}
\caption{Transmission probability versus the energy $E$ in $K$ and $K'$ valleys for  $L=10$ nm, $k_y=0.5~\text{nm}^{-1}$, and four magnetic field values  Panels (a): $B= 0.1$ T, (b): $B= 0.2$ T, (c): $B= 0.3$ T, and (d): $B= 0.4$ T.}\label{T3}
\end{figure}

\begin{figure}[htp!]
\includegraphics[scale=0.4]{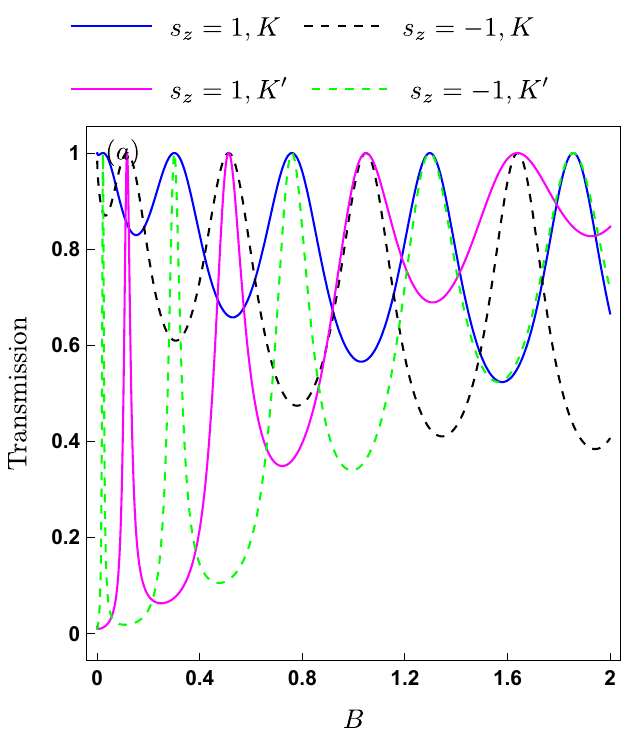}\includegraphics[scale=0.4]{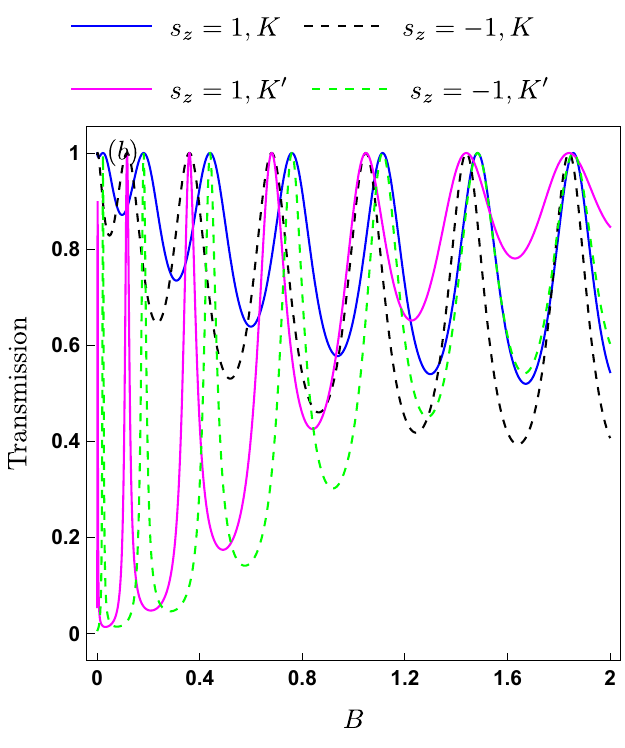}\\
\includegraphics[scale=0.4]{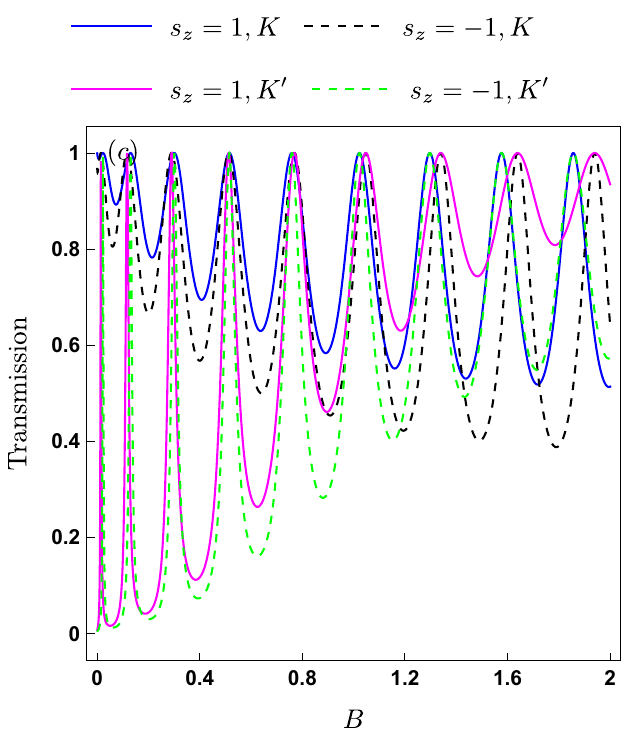}\includegraphics[scale=0.4]{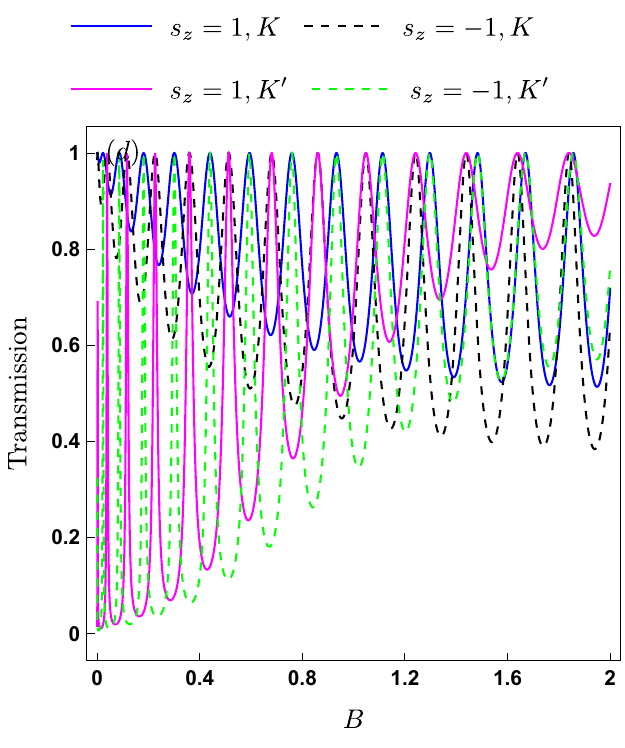}
\caption{Transmission probability versus the magnetic field $B$ in  $K$ and $K'$ valleys for  $E=2.$ eV,  $k_y=0.5~\text{nm}^{-1}$, and barrier width values (a): $L= 10$ nm, (b): $L= 15$ nm, (c): $L= 20$ nm, (d): $L= 30$ nm.
}\label{T4}
\end{figure}

Fig. \ref{T3} shows the transmission probability as a function of electron energy $E$ for electrons in the $K$ and $K'$ valleys for $L = 10$ nm, $k_y = 0.5~\text{nm}^{-1}$, and $B = 0.1$, 0.2, 0.3, 0.4 T in (a), (b), (c), (d), respectively. As the electron energy increases beyond the band gap, the transmission also increases and exhibits several resonance peaks. These peaks are due to the formation of quasi-bound states inside the barrier, where constructive interference occurs at certain energies. At the lower field of $B = 0.1$ T in  Fig. \ref{T3}(a), only a few resonance peaks are visible. As the magnetic field increases, the magnetic confinement becomes stronger and the resonance peaks become sharper and more distinct. The resonance patterns also differ between the $K$ and $K'$ valleys. The positions and heights of the peaks vary, showing that the spin-orbit coupling causes valley-dependent changes in the effective potential. The valley differences become larger at higher magnetic fields. This shows that both magnetic effects and spin-orbit coupling influence the electron flow in MoS$_2$. Our results are in agreement with the work \cite{hao2020influence}, which found similar changes in the carrier velocity. Together, they prove that the bandgap and spin-orbit coupling of MoS$_2$ create separate spin-valley paths and multiple resonances.

Figure~\ref{T4} shows the transmission probability versus the magnetic field $B$ in the $K$ and $K'$ valleys for $E = 2$ eV, $k_y = 0.5~\text{nm}^{-1}$, and $L = 10$, 15, 20, and 30 nm in (a), (b), (c), (d) , respectively. The transmission curves show oscillations that are Fabry-Perot type resonances. They are caused by repeated reflections of the electron wave at the barrier edges. As the barrier width increases, the interference becomes more sensitive to changes in the magnetic field. This leads to noticeable changes in both the strength and spacing of the resonance peaks. The $K$ and $K'$ valleys also show different resonance patterns. This difference is caused by spin-orbit coupling, which affects how the electrons in each valley interfere.


\section{Conductance}\label{CCCC}
\begin{figure}[htp!]
\centering 
\includegraphics[scale=0.55]{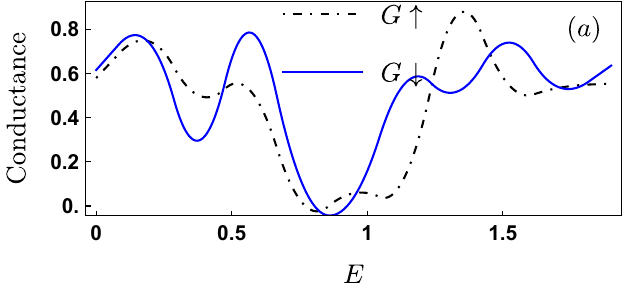}\\
\includegraphics[scale=0.55]{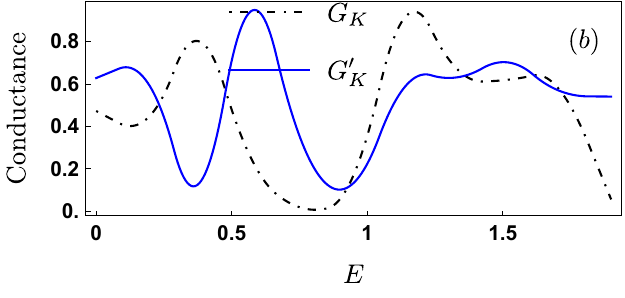}
\caption{Spin- and valley-dependent conductance  {$G_{\uparrow (\downarrow)}\equiv G_{\uparrow (\downarrow)}/G_0$} (a) and 
{$G_{K (K')}  \equiv G_{K (K')}/G_0$} (b) versus the energy $E$ for $B=0.2$ T, and $L=25$ nm. 
}\label{Conductance}
\end{figure}

The conductance is an important measure for evaluating the performance of spin-valley devices, as it indicates how easily charge carriers move through the system under external influences. In this study, we define the conductance for a given valley $\tau$ and spin $s$ by integrating the transmission probability over all possible incident angles. 
The conductance is expressed as \cite{buttiker1986four}
\begin{align}\label{GGGG}
 G_{\tau, s} = G_0 \int_{-\pi/2}^{\pi/2} T_{\tau, s}(\theta) \cos\theta \, d\theta,
\end{align}
 where \( T_{\tau, s}(\theta) \) is the transmission probability for electrons with valley index \( \tau \) and spin index \( s \), depending on the incident angle \( \theta \). The factor \( G_0 \) represents the conductance quantum scaled by system-specific constants. The evaluation of  \eqref{GGGG} can help us understand how spin, valley, and magnetic field affect electron transport. As a result, we can gain useful insight into how spin- and valley-polarized currents behave in monolayer MoS$_2$. For numerical use, we introduce 
 the averaged conductances as
 \begin{align}
 	&G_{\uparrow(\downarrow)} = \frac{G_{K\uparrow(\downarrow)} + G_{K'\uparrow(\downarrow)}}{2},\\
 	&G_{K(K')} = \frac{G_{K(K')\uparrow} + G_{K(K')\downarrow}}{2}.
 \end{align}
 
 Fig. \ref{Conductance} presents the spin- and valley-dependent conductance versus the energy for $B = 0.2$ T and $L = 25$ nm. The conductance is calculated separately for the spin-up and spin-down channels ($G_{\uparrow},G_{\downarrow}$) as well as for the $K$ and $K'$ valleys ($G_{K},G_{K'}$). As the energy increases beyond the intrinsic band gap, the conductance also increases and shows oscillations. These oscillations are a sign of the formation of quasi-bound states inside the barrier. At lower energies, there are noticeable differences between the spin-up and spin-down conductances due to spin-orbit coupling. We observe a clear separation in conductance between the $K$ and $K'$ valleys, indicating that the transport is valley dependent. This is in agreement with previous studies on MoS$_2$ \cite{hao2020influence}, which showed that tuning the Fermi velocity or applying a gate voltage can switch between different spin and valley states. As a result, spin- and valley-polarized currents can be precisely and selectively controlled.

\section{MoS$_2$ and  graphene}\label{MoGr}

	Electrons move through MoS$_2$ and graphene in very different ways at the quantum level. In MoS$_2$, whether an electron can pass through a barrier depends strongly on whether it is coming from the left or right (Fig.~\ref{T0}). This directional behavior comes from a special property called spin-valley locking, in which   spin, motion, and valley type of the electron are closely linked \cite{xiao2012coupled}. Graphene, however, behaves quite differently. Its electrons can easily pass through barriers  from any direction, thanks to a phenomenon known as Klein tunneling~\cite{katsnelson2006chiral}. Things get even more interesting in MoS$_2$ when we look at electrons from the two valleys ($K$ and $K'$). Their ability to pass through changes depending on their sideways motion ($k_y$), as shown in Fig.~\ref{T1}. In graphene, by contrast, both valleys act the same and contribute equally to how current flows~\cite{geim2007rise}.
Unlike graphene, which allows electrons to pass through with an almost full  transmission probability at any energy thanks to its massless Dirac fermions~\cite{geim2009graphene}, MoS$_2$ behaves very differently. As shown in Fig.~\ref{T3}, MoS$_2$ blocks all transmission below its 1.8 eV bandgap~\cite{mak2010atomically}, and only allows electrons through once they have sufficient energy. This often results in sharp, resonant peaks just above the threshold. Interestingly,  transmission of MoS$_2$ does not simply increase with the magnetic field, it also  oscillates in a non-monotonic way (see Fig.~\ref{T4}). These oscillations result from a complex interplay between Landau level formation and spin-valley splitting~\cite{PhysRevLett.120.057405}—a behavior that graphene cannot  reproduce, since it lacks both a bandgap and strong spin-orbit coupling~\cite{Min2006}. These unique properties make MoS$_2$ a much better candidate for applications requiring control over electron direction, valley identity, or magnetic tuning. These are areas where the full transmission of graphene limits its use.

The conductance behavior of monolayer MoS$_2$ under a magnetic barrier exhibits pronounced spin-- and valley--dependent characteristics, setting it apart fundamentally from graphene. Unlike graphene, which---despite its unique Landau-level spectrum in magnetic fields---retains spin and valley degeneracy in transport due to negligible intrinsic spin--orbit coupling (SOC) and symmetrical Dirac cones \cite{Neto2009,Novoselov2012}, MoS$_2$ features a finite bandgap (\~1.8\,eV), strong SOC, and broken inversion symmetry, leading to distinct transport channels for spin-up/spin-down and $K/K'$ valleys. Our results (Fig.~\ref{Conductance}) show clear energy-dependent separation in conductance among these channels, especially at low energies where SOC causes noticeable spin asymmetry. As energy increases beyond the intrinsic gap, conductance rises with pronounced oscillations---signatures of quasi-bound states within the magnetic barrier. In graphene, magnetic barriers induce similar oscillations but without spin/valley selectivity, as its relativistic carriers experience only pseudospin-dependent scattering \cite{Neto2009,Masir2013}.
While graphene's lack of a bandgap enables current flow at all energies \cite{Neto2009}, MoS$_2$ acts as a switch, turning on only when the energy exceeds its threshold. This dichotomy is critical for transistor applications, where gate-controlled on/off states are essential. Moreover, MoS$_2$'s strong SOC and valley contrast allow selective control of spin and valley channels via Fermi-level tuning \cite{xiao2012coupled}, whereas graphene requires extrinsic modifications (e.g., proximity coupling, strain, or heterostructures \cite{Schaibley2016}) to break degeneracies. Thus, while graphene excels in revealing fundamental quantum phenomena like Klein tunneling or ultra-relativistic Landau quantization \cite{Neto2009,Novoselov2012}, MoS$_2$ offers intrinsic tunability for functional spin/valleytronic devices.

One feasible approach to experimental validation is to perform low-temperature magnetotransport measurements on monolayer MoS$_2$ under inhomogeneous magnetic fields. This can be achieved using ferromagnetic stripes or patterned gates. These setups can mimic the magnetic barrier structures modeled in our work and those successfully implemented in graphene-based systems \cite{Masir2013}. Additionally, scanning tunneling microscopy (STM) and spectroscopy (STS) could probe the local density of states and edge-resolved transport features in MoS$_2$ flakes subjected to spatially modulated magnetic fields. Recent STM studies on 2D materials show promise in resolving spin- and valley-polarized states at the nanometer scale \cite{Zhang2014}. Finally, nonlocal transport measurements and valley Hall setups, as demonstrated in recent MoS$_2$ experiments, offer another method for detecting valley-polarized currents induced by external fields \cite{mak2014valley}. These methods could reveal the conductance modulation and valley-dependent features predicted by our model.

\section{Conclusion}\label{S4}

We have studied the electron transport properties of monolayer MoS$_2$ with a magnetic barrier using a full-band continuum model. Our results show that magnetic confinement significantly affects electron transport through the formation of resonant tunneling channels. When the electron energy exceeds the intrinsic bandgap, quasi-bound states arise in the barrier region due to quantum interference between forward and backward propagating electron waves. These states give rise to distinct resonance peaks in the transmission spectrum. The characteristics of these resonant features are highly sensitive to changes in both the magnetic field strength and the width of the barrier. As the magnetic field increases, so does the effective wave vector inside the barrier. This changes the phase accumulation conditions, resulting in a shift of the resonance positions and leading to sharper and more pronounced transmission peaks. In addition, the interplay between magnetic confinement and spin-orbit coupling introduces spin- and valley-dependent effects that further modulate the transmission behavior. These variations manifest as asymmetric transmission profiles for the $K$ and $K'$ valleys and for spin-up and spin-down states, especially under strong magnetic fields.

By analyzing the transmission, reflection probabilities, and conductance, we show that spin and valley channels can be selectively enhanced or suppressed by tuning external parameters such as magnetic field strength, electron energy, barrier width, and transverse wave vector. In particular, the conductance study reveals clear spin and valley polarization effects, consistent with the emergence of spin-valley locking induced by the strong spin-orbit interaction in MoS$_2$. The results also confirm that Fabry-Perot type interference is responsible for the oscillatory behavior observed in conductance and transmission, especially at energies above the band gap. These results highlight the potential of monolayer MoS$_2$ as a platform for controllable spintronic and valleytronic devices. 

The ability to manipulate quantum transport through external magnetic barriers opens new avenues for the design of nanoscale devices that exploit the internal degrees of freedom of electrons, namely spin and valley indices, for information processing and storage. Future work could explore the influence of additional factors such as strain, disorder, and electric field modulation to extend the tunability and performance of such systems in realistic device architectures.

As for the future outlook, our work can be extended in several promising directions.
First, we can study temperature effects. We focused on the zero-temperature regime to capture intrinsic, coherent spin- and valley-dependent transport without thermal broadening. At low temperatures ($\leq$ 30 K), key features such as sharp resonances and polarization should remain observable, as supported by experimental results \cite{Yang2015}. At higher temperatures, increased phonon scattering and Fermi broadening are expected to reduce polarization and smooth transmission features. Second, regarding disorder, we assumed an ideal, defect-free MoS$_2$ monolayer. However, real samples inevitably contain imperfections. These imperfections can lead to resonance broadening and reduced spin/valley polarization. Nevertheless, the strong spin–orbit coupling and large valley separation MoS$_2$ provide robustness against moderate disorder \cite{Zhou2013}. Third, extending the model to periodic or stacked magnetic barriers (e.g., magnetic superlattices) could enhance spin/valley filtering and lead to miniband formation and tunable transport gaps, as seen in graphene \cite{Masir2013}. These effects may be stronger in MoS$_2$ due to its intrinsic spin-valley coupling. Fourth, incorporating gate-tunable electric fields alongside magnetic barriers in hybrid structures could enable dynamic control of band profiles and further enhance spin and valley selectivity.

\section*{Acknowledgment}
P.D. and D.L. acknowledge partial financial support from FONDECYT 1231020.


\begin{thebibliography}{99}
	
	\bibitem{xia2014two}
	F.~Xia, H.~Wang, D.~Xiao, M.~Dubey, and A.~Ramasubramaniam,
	``Two-dimensional material nanophotonics,''
	\textit{Nat. Photon.} \textbf{8}, 899--907 (2014).
	
	\bibitem{liu2019van}
	Y.~Liu, Y.~Huang, and X.~Duan,
	``Van der Waals integration before and beyond two-dimensional materials,''
	\textit{Nature} \textbf{567}, 323--333 (2019).
	
	\bibitem{chowdhury2020progress}
	T.~Chowdhury, E.~C. Sadler, and T.~J. Kempa,
	``Progress and prospects in transition-metal dichalcogenide research beyond 2D,''
	\textit{Chem. Rev.} \textbf{120}, 12563--12591 (2020).
	
	\bibitem{fu20212d}
	Q.~Fu \textit{et al.},
	``2D transition metal dichalcogenides: design, modulation, and challenges in electrocatalysis,''
	\textit{Adv. Mater.} \textbf{33}, 1907818 (2021).
	
	\bibitem{yang20232d}
	R.~Yang \textit{et al.},
	``2D transition metal dichalcogenides for photocatalysis,''
	\textit{Angew. Chem. Int. Ed.} \textbf{135}, e202218016 (2023).
	
	\bibitem{yun2020layered}
	Q.~Yun \textit{et al.},
	``Layered transition metal dichalcogenide-based nanomaterials for electrochemical energy storage,''
	\textit{Adv. Mater.} \textbf{32}, 1903826 (2020).
	
	\bibitem{lv2015transition}
	R.~Lv \textit{et al.},
	``Transition metal dichalcogenides and beyond: synthesis, properties, and applications of single-and few-layer nanosheets,''
	\textit{Acc. Chem. Res.} \textbf{48}, 56--64 (2015).
	
	\bibitem{gatensby2014controlled}
	R.~Gatensby \textit{et al.},
	``Controlled synthesis of transition metal dichalcogenide thin films for electronic applications,''
	\textit{Appl. Surf. Sci.} \textbf{297}, 139--146 (2014).
	
	\bibitem{lembke2015single}
	D.~Lembke, S.~Bertolazzi, and A.~Kis,
	``Single-layer MoS2 electronics,''
	\textit{Acc. Chem. Res.} \textbf{48}, 100--110 (2015).
	
	\bibitem{gong2017electronic}
	C.~Gong \textit{et al.},
	``Electronic and optoelectronic applications based on 2D novel anisotropic transition metal dichalcogenides,''
	\textit{Adv. Sci.} \textbf{4}, 1700231 (2017).
	
	\bibitem{pospischil2016optoelectronic}
	A.~Pospischil and T.~Mueller,
	``Optoelectronic devices based on atomically thin transition metal dichalcogenides,''
	\textit{Appl. Sci.} \textbf{6}, 78 (2016).
	
	\bibitem{choi2017recent}
	W.~Choi \textit{et al.},
	``Recent development of two-dimensional transition metal dichalcogenides and their applications,''
	\textit{Mater. Today} \textbf{20}, 116--130 (2017).
	
	\bibitem{ahn20202d}
	E.~C. Ahn,
	``2D materials for spintronic devices,''
	\textit{npj 2D Mater. Appl.} \textbf{4}, 17 (2020).
	
	\bibitem{feng2017prospects}
	Y.~P. Feng \textit{et al.},
	``Prospects of spintronics based on 2D materials,''
	\textit{WIREs Comput. Mol. Sci.} \textbf{7}, e1313 (2017).
	
	\bibitem{li2016two}
	X.~Li and X.~Wu,
	``Two-dimensional monolayer designs for spintronics applications,''
	\textit{WIREs Comput. Mol. Sci.} \textbf{6}, 441--455 (2016).
	
	\bibitem{geim2007rise}
	A.~K. Geim and K.~S. Novoselov,
	``The rise of graphene,''
	\textit{Nat. Mater.} \textbf{6}, 183--191 (2007).
	
	\bibitem{geim2009graphene}
	A.~K. Geim,
	``Graphene: status and prospects,''
	\textit{Science} \textbf{324}, 1530--1534 (2009).
	
	\bibitem{cooper2012experimental}
	D.~R. Cooper \textit{et al.},
	``Experimental review of graphene,''
	\textit{ISRN Nanotechnol.} \textbf{2012}, 501686 (2012).
	
	\bibitem{splendiani2010emerging}
	A.~Splendiani \textit{et al.},
	``Emerging photoluminescence in monolayer MoS2,''
	\textit{Nano Lett.} \textbf{10}, 1271--1275 (2010).
	
	\bibitem{mak2010atomically}
	K.~F. Mak \textit{et al.},
	``Atomically thin MoS2: a new direct-gap semiconductor,''
	\textit{Phys. Rev. Lett.} \textbf{105}, 136805 (2010).
	
	\bibitem{xiao2012coupled}
	D.~Xiao, G.-B. Liu, W.~Feng, X.~Xu, and W.~Yao,
	``Coupled spin and valley physics in monolayers of MoS2 and other group-VI dichalcogenides,''
	\textit{Phys. Rev. Lett.} \textbf{108}, 196802 (2012).
	





	\bibitem{wang2012large}
	H.~Wang \textit{et al.},
	``Large-scale 2D electronics based on single-layer MoS2 grown by chemical vapor deposition,''
	in \textit{IEDM Tech. Dig.}, 4--6 (IEEE, 2012).
	
	\bibitem{wang2012integrated}
	H.~Wang \textit{et al.},
	``Integrated circuits based on bilayer MoS2 transistors,''
	\textit{Nano Lett.} \textbf{12}, 4674--4680 (2012).
	
	\bibitem{lee2012synthesis}
	Y.-H. Lee \textit{et al.},
	``Synthesis of large-area MoS2 atomic layers with chemical vapor deposition,''
	\textit{arXiv:1202.5458} (2012).
	
	\bibitem{liu2012growth}
	K.-K. Liu \textit{et al.},
	``Growth of large-area and highly crystalline MoS2 thin layers on insulating substrates,''
	\textit{Nano Lett.} \textbf{12}, 1538--1544 (2012).
	
	\bibitem{zhan2011large}
	Y.~Zhan, Z.~Liu, S.~Najmaei, P.~M. Ajayan, and J.~Lou,
	``Large area vapor phase growth and characterization of MoS2 atomic layers on SiO2 substrate,''
	\textit{arXiv:1111.5072} (2011).
	
	\bibitem{radisavljevic2011single}
	B.~Radisavljevic \textit{et al.},
	``Single-layer MoS2 transistors,''
	\textit{Nat. Nanotechnol.} \textbf{6}, 147--150 (2011).
	
	\bibitem{kaasbjerg2012phonon}
	K.~Kaasbjerg, K.~S. Thygesen, and K.~W. Jacobsen,
	``Phonon-limited mobility in n-type single-layer MoS2 from first principles,''
	\textit{Phys. Rev. B} \textbf{85}, 115317 (2012).
	
	\bibitem{cheianov2006selective}
	V.~V. Cheianov and V.~I. Fal'ko,
	``Selective transmission of Dirac electrons and ballistic magnetoresistance of n-p junctions in graphene,''
	\textit{Phys. Rev. B} \textbf{74}, 041403 (2006).
	
	\bibitem{katsnelson2006chiral}
	M.~I. Katsnelson, K.~S. Novoselov, and A.~K. Geim,
	``Chiral tunnelling and the Klein paradox in graphene,''
	\textit{Nat. Phys.} \textbf{2}, 620--625 (2006).
	
	\bibitem{mak2019probing}
	K.~F. Mak, J.~Shan, and D.~C. Ralph,
	``Probing and controlling magnetic states in 2D layered magnetic materials,''
	\textit{Nat. Rev. Phys.} \textbf{1}, 646--661 (2019).
	
	\bibitem{premasiri2019tuning}
	K.~Premasiri and X.~P. Gao,
	``Tuning spin-orbit coupling in 2D materials for spintronics: a topical review,''
	\textit{J. Phys.: Condens. Matter} \textbf{31}, 193001 (2019).
	
	
	
	\bibitem{nogaret2000resistance}
	A.~Nogaret, S.~J. Bending, and M.~Henini,
	``Resistance resonance effects through magnetic edge states,''
	\textit{Phys. Rev. Lett.} \textbf{84}, 2231 (2000).
	
	\bibitem{papp2001spin}
	G.~Papp and F.~M. Peeters,
	``Spin filtering in a magnetic-electric barrier structure,''
	\textit{Appl. Phys. Lett.} \textbf{78}, 2184--2186 (2001).
	
	\bibitem{li2013unconventional}
	X.~Li, F.~Zhang, and Q.~Niu,
	``Unconventional quantum Hall effect and tunable spin Hall effect in Dirac materials: application to an isolated MoS2 trilayer,''
	\textit{Phys. Rev. Lett.} \textbf{110}, 066803 (2013).
	
	
		\bibitem{zahid2013generic}
	F.~Zahid, L.~Liu, Y.~Zhu, J.~Wang, and H.~Guo,
	``A generic tight-binding model for monolayer, bilayer and bulk MoS2,''
	\textit{AIP Adv.} \textbf{3}, 052123 (2013).
	
	
	
	
	\bibitem{cheng2015transport}
	F.~Cheng, Y.~Ren, and J.-F. Sun,
	``Transport through a Single Barrier on Monolayer MoS2,''
	\textit{Chin. Phys. Lett.} \textbf{32}, 107301 (2015).
	

	\bibitem{mak2014valley}
	K.~F. Mak, K.~L. McGill, J.~Park, and P.~L. McEuen,
	``The valley Hall effect in MoS2 transistors,''
	\textit{Science} \textbf{344}, 1489--1492 (2014).
	
	
	
	
	\bibitem{hao2020influence}
	X.-J. Hao, R.-Y. Yuan, J.-J. Jin, and Y.~Guo,
	``Influence of the velocity barrier on the massive Dirac electron transport in a monolayer MoS2 quantum structure,''
	\textit{Front. Phys.} \textbf{15}, 1--9 (2020).
	
	
		\bibitem{buttiker1986four}
	M.~B\"uttiker,
	``Four-terminal phase-coherent conductance,''
	\textit{Phys. Rev. Lett.} \textbf{57}, 1761 (1986).
	
	
	\bibitem{PhysRevLett.120.057405}
	A.~V. Stier, N.~P. Wilson, K.~A. Velizhanin, J.~Kono, X.~Xu, and S.~A. Crooker,
	``Magnetooptics of Exciton Rydberg States in a Monolayer Semiconductor,''
	\textit{Phys. Rev. Lett.} \textbf{120}, 057405 (2018).
	
	
	\bibitem{Min2006}
	H.~Min, J.~E. Hill, N.~A. Sinitsyn, B.~R. Sahu, L.~Kleinman, and A.~H. MacDonald,
	``Intrinsic and Rashba spin-orbit interactions in graphene sheets,''
	\textit{Phys. Rev. B} \textbf{74}, 165310 (2006).
	
	
		\bibitem{Neto2009}
	A.~H. Castro Neto, F.~Guinea, N.~M. R. Peres, K.~S. Novoselov, and A.~K. Geim,
	``The electronic properties of graphene,''
	\textit{Rev. Mod. Phys.} \textbf{81}, 109--162 (2009).
	
	\bibitem{Novoselov2012}
	K.~S. Novoselov, V.~I. Fal'ko, L.~Colombo, P.~R. Gellert, M.~G. Schwab, and K.~Kim,
	``A roadmap for graphene,''
	\textit{Nature} \textbf{490}, 192--200 (2012).
	
	\bibitem{Masir2013}
	M.~R. Masir, P.~Vasilopoulos, and F.~M. Peeters,
	``Magnetic barrier structures in graphene,''
	\textit{Phys. Rev. B} \textbf{88}, 245413 (2013).
	
	\bibitem{Schaibley2016}
	J.~R. Schaibley, H.~Yu, G.~Clark, P.~Rivera, J.~S. Ross, K.~L. Seyler, W.~Yao, and X.~Xu,
	``Valleytronics in 2D materials,''
	\textit{Nat. Rev. Mater.} \textbf{1}, 16055 (2016).
	
	
	
	\bibitem{Zhang2014}
	C.~Zhang, A.~Johnson, C.-L. Hsu, L.-J. Li, and C.-K. Shih,
	``Direct Imaging of Band Profile in Single Layer MoS$_2$ on Graphite: Quasiparticle Energy Gap, Metallic Edge States, and Edge Band Bending,''
	\textit{Nano Lett.} \textbf{14}, 2443--2447 (2014).
	
	
	\bibitem{Yang2015}
	L.~Yang, N.~A. Sinitsyn, W.~Chen, J.~Yuan, J.~Zhang, J.~Lou, and S.~A. Crooker,
	``Long-lived nanosecond spin relaxation and spin coherence of electrons in monolayer MoS$_2$ and WS$_2$,''
	\textit{Nat. Phys.} \textbf{11}, 830--834 (2015).
	
	\bibitem{Zhou2013}
	W.~Zhou, X.~Zou, S.~Najmaei, Z.~Liu, Y.~Shi, J.~Kong, J.~Lou, P.~M. Ajayan, B.~I. Yakobson, and J.-C. Idrobo,
	``Intrinsic structural defects in monolayer molybdenum disulfide,''
	\textit{Nano Lett.} \textbf{13}, 2615--2622 (2013).
	
	
%
%
%
%
%
%
%
%
%
%
%
	
	
	
	
	
	
\end{thebibliography}
\end{document}